\documentclass[10pt, a4paper,twocolumn]{article}
\usepackage[utf8]{inputenc}
\usepackage[T1]{fontenc}

\usepackage{amsmath}
\usepackage{amsfonts}
\usepackage{subcaption}
\usepackage{nicefrac}
\usepackage{authblk}
\usepackage{graphicx}

\usepackage[left=2.5cm,right=2.5cm,top=3cm,bottom=3cm,bindingoffset=0cm]{geometry}

\begin{document}

\title{\bf Data Optimisation for a Deep Learning Recommender System}

\author[1]{Gustav Hertz} %\thanks{gustav.hertz@ingka.ikea.com}}

\author[1]{Sandhya Sachidanandan} %\thanks{sandhya.sachidanandan@ingka.ikea.com}}
\author[1]{Bal\'{a}zs T\'{o}th} %\thanks{balazs.toth@ingka.ikea.com}}

\author[1]{\\ Emil S. J\o rgensen} %\thanks{emil.joergensen@ingka.ikea.com}}
\author[1,2]{Martin Tegn\'{e}r} %\thanks{Corresponding author.}}

% \affil[  ]{\texttt{\{gustav.hertz, sandhya.sachidanandan, balazs.toth, emil.joergensen, martin.tegner\}@ingka.ikea.com}}
% \affil[  ]{}

\affil[1]{IKEA Group: \texttt{\{gustav.hertz, sandhya.sachidanandan, balazs.toth, emil.joergensen, martin.tegner\}@ingka.ikea.com}}
\affil[2]{Oxford-Man Institute, University of Oxford}

\date{}

\newcommand{\vect}[1]{{#1}}

\maketitle

\begin{abstract}
This paper advocates privacy preserving requirements on collection of user data for recommender systems.  The purpose of our study is twofold. First, we ask if restrictions on data collection will hurt test quality of RNN-based recommendations.  We study how validation performance depends on the available amount of training data. We use a combination of top-K accuracy, catalog coverage and novelty for this purpose, since good recommendations for the user is not necessarily captured by a traditional accuracy metric. Second, we ask if we can improve the quality under minimal data by using secondary data sources. We propose knowledge transfer for this purpose and construct a representation to measure similarities between purchase behaviour in data. This to make qualified judgements of which source domain will contribute the most. Our results show that (i) there is a saturation in test performance when training size is increased above a critical point. We also discuss the interplay between different performance metrics, and properties of data. Moreover, we demonstrate that (ii) our representation is meaningful for measuring purchase behaviour. In particular, results show that  we can leverage secondary data to improve validation performance if we select a relevant source domain according to our similarly measure.

\end{abstract}

%\keywords{datasets, LSTM, privacy, recommender systems}

\maketitle

\section{Introduction}
In the last few years, considerations of  \textit{user privacy} and  \textit{data security} have gained attention in the AI community along with  principles such as \textit{algorithmic fairness}  and \textit{bias} (see e.g. \cite{al2019privacy, mehrabi2019survey}). 
This is relevant for learning algorithms that make decisions based on data from users: movie recommendations, product suggestions, loan applications and match-making sites are common applications, to mention just but a few. Indeed, the  2012 White House report on privacy and consumer data \cite{house2012consumer}, and the recent EU General Data Project Regulation in 2018, deem  considerations of privacy and security  unavoidable for the deployer of user-centric systems.

On the other side of the coin is the fact that progress in deep learning is partially enabled by large training sets of representative data. Neural networks have the potential and  a proven record of modelling input-to-output mappings with unlimited complexity and flexibility (\cite{he2015delving, silver2016mastering, vinyals2015grammar} are some examples), but they are \textit{data greedy}, especially to generalise well to unseen test data. ``More is more'' is thus the reign in the development phase of deep learning systems,  typically on a centralised repository of  collected data. In many applications, such as image classification and machine translation, this is  not an issue, while  interaction with users requires care in both the collection and storage of their data.

This paper considers \textit{data optimisation} for a deep-learning recommender system. First, we study how the recommender system’s {performance} on validation data depends on the \textit{size} of the training data. To this end, we use a set of performance metrics that are designed to measure the quality of a recommendation. This since `good' recommendation for the user is not necessarily  represented by an observable output variable; we do not have a ground-truth as our primary target,\footnote{This in contrast to e.g. image classification, where we have access to a true output label (`cat', `dog' etc.) associated with the input image.} but rather heuristic measures for the success of a recommendation. { From experiments, we conclude that there is an optimal amount of training data beyond which the performance of the model either saturates or decreases. We discuss this in terms of properties of both the metrics and the generating data distribution}. Second, we study how we can improve the performance under a minimal data requirement. We use  knowledge transfer  for this purpose under the assumption that we can leverage  data from secondary sources. %other, possibly heterogeneous, but related domains.
{This since our study is based on a multi-market setting.} We propose a representation of purchase behaviour for a similarity measure to judge which secondary data distribution is most suitable for our task at hand.  We study the effectiveness of the knowledge transfer and show that we can achieve significant performance gains in the target domain by leveraging a source domain according to our similarity measure.

In the next section, we discuss the motivation and background of our work along with related literature. We describe our setup in Section \ref{sec:expriment}, including purchase data, recommender algorithm, performance metrics and construction behind the representation for our similarity measure. Section \ref{sec:expRes} describes the experiments and discusses their results. Section \ref{sec:conclusion} summaries and concludes our work. %We  describe the purchase data that is used in our work. We establish some notation, and detail the deep learning model at the base of the recommender system along with the metrics we use for quantifying its performance. The construction of our representation of purchase behaviour and the similarity measure is described at the end of the section. 

\section{Background and Related Work}
The main motivation behind our work is to put data requirements along with user privacy at the heart of the development and deployment of AI-based systems for recommendations. %We do this through a notion of \textit{minimal necessary data} and argue for an ex-post analysis of limited {collection} rather than ex-ante {removal} of user data.
%We support our case with empirical evidence that recommendation performance \textit{saturate} with amount of data. Our proposition is to use techniques from transfer learning to  ...

\paragraph{Minimal necessary data}
In a recent paper \cite{larson2017towards}, the authors advocate a set of best-practice principles of \textit{minimal necessary data} and \textit{training-data requirements analysis}.  We follow their motivation, which builds on the assumption that data-greed is also a (bad) habit in the development of recommender systems. The authors emphasise that data from users is a \textit{liability}; respect for user privacy should be considered during data collection, and the data itself protected and securely stored. At the same time, they state that there is a clear notion of performance \textit{saturation} in
the square-error 
 metric when the amount of training data reaches a certain point. Hence, it is not only a desire in view of user privacy to analyse requirements on training data, but also highly sensible: collecting more than necessary data should always be discouraged, and the trade-off between marginal performance to additional data  act as guiding principle.

A similar motivation is central in \cite{chow2013differential}. The authors suggests a \textit{differential data analysis} for understanding which data contributes to performance in recommender systems, and propose that less useful data should be discarded based on the analysis. While we fully share their motivation and view that performance saturates with data size (as empirically confirmed in \cite{larson2017towards}), we like to highlight the \textit{post-hoc} nature of their analysis. The choice of which particular data should be collected and eventually discarded is made after the data has been analysed. In particular, it is in the hands of the system owner. %, not the user. 
If the control of data collection is shifted to the user itself, such tools for  \textit{removal} of data are less powerful.%\footnote{This notion of privacy is  associated with data as a \textit{tangible} asset, with focus on secure storage and also the  location of data. \textit{Decentralisation} \cite{mcmahan2017communication} is a recent method that circumvents the liability of centralised data, by a  distribution of the algorithm to use data only locally on-device. Similarly, the idea of \textit{differential privacy} is incorporated in deep learning in \cite{abadi2016deep} to protect users from adversarial attacks that could lead to retrieval of sensitive data. We highlight privacy already at the level of  \textit{usage} of data. If,  due to privacy considerations, ownership of this control is given to  users, there is probably much less data to collect in the  first place.}

While we put user privacy and \textit{ex-ante} minimal collection of data as our primary motivator, there is also studies on how algorithms perform when the amount of training data is naturally limited, see e.g. \cite{forman2004learning}. For recommender systems, this is the cold-start problem and \cite{cremonesi2009analysis} analyse how algorithms compare in this situation. % along with their dependency on size of training data. 
Another approach to circumvent limitations on data is to use contextual information about items as basis for the recommender system in place of sensitive user-information. \cite{mt20prob} propose an approach where the system’s interaction with user data can be minimal, at least when there is strong \textit{a-priori} knowledge about how items relate in a relevant way to recommendations.

\paragraph{Differential privacy and federated learning}
As in \cite{chow2013differential}, the notion of privacy is commonly associated with data as a \textit{tangible} asset, with focus on secure storage, distribution and also  location of user data.  \textit{Decentralisation} \cite{mcmahan2017communication} is a recent method that circumvents  liabilities of centralising the data, by a  distribution of the algorithm to use data only locally on-device. Similarly, the idea of \textit{differential privacy} is incorporated in deep learning in \cite{abadi2016deep} to protect users from adversarial attacks that could lead to retrieval of sensitive data. While this stream of research is equally important, we highlight privacy concerns already at the level of \textit{usage} of data. If,  for privacy considerations, ownership of the control of usage is given to users, there is probably much less data to collect in the very first place.

\paragraph{Meaningful recommendations} 
%\textit{Discuss why and how we need to consider different measures, to measure ``performance'' of the system. Discuss what is `good' for the user. This in place of 'traditional' metrics, such as accuracy, precision, recall etc. Consider references [7, 10] in Larson et al.}
In the context of recommendations, it is not clear that common metrics of accuracy represents performance in a way that is meaningful for the user, see e.g. \cite{mcnee2006being}. First, there is no obvious target that is directly representative for  performance of an algorithm. Second, since recommender systems are {feedback} systems, they should ultimately be validated in an online manner, see \cite{10.1145/2487575.2488215, 10.1145/2645710.2645745, 10.1145/2792838.2800184}. However, combining accuracy with offline diversity metrics like coverage and novelty can help the process of optimizing recommender systems for online performance, \cite{10.1145/2792838.2800184}.  We therefore use metrics that are heuristically constructed to measure performance in recommendations in terms of quality for the user: top-k accuracy, novelty and coverage  
\cite{beyondacc, nov_and_div}.

%  {\color{blue} \cite{larson2017towards} describes a set of classic recommender system experiments where the trade-off between training data volume and Root Mean Squared Error (RMSE) is examined. We make further experiments about the trade-off between training data volume and some metrics that are commonly used for offline evaluation of recommender systems. These metrics can give a good indication about the online performance of the recommender, and online performance is what most people care about. One of the metrics we use is top-k accuracy. Although, it is a good indicator of the performance of recommender models, it is not perfect. Accuracy metrics don't always correlate with online metrics like click-through rate (CTR) \cite{10.1145/2487575.2488215, 10.1145/2645710.2645745}. Therefore, we use two additional metrics that are advised by : novelty and coverage. Combining accuracy metrics with other metrics like coverage and novelty can help the process of optimizing recommender models for online metrics \cite{10.1145/2792838.2800184}.
% }
 %
%\textit{Discuss other ways of 'solving' issues around privacy: Differential privacy does not attack the core---use less data---but rather it's a suggestion to protect data from adversarial attacks. Similarly, federated learning is a 'hands-off' approach, to avoid collection and storage, while it still 'uses' the same data (hence not ``good enough'' for IKEA's data promise).}

\paragraph{Knowledge transfer} 
For recommendations, it is common practice to use transfer learning  in collaborative algorithms, see e.g. \cite{Pan2011TransferLT, pan2010transfer, li2009can} and the survey \cite{DBLP:journals/ijon/Pan16}.  While these approaches typically address data sparsity and missing ratings for collaborative recommendations, %with \textit{transductive} methods, 
we focus on improving performance under limited amounts of training data in the target domain by \textit{inductive} instance-based learning (see \cite{PanY09TKDE}). We consider a multi-market setup, and hypothesise that we can make an informed selection of source data for our knowledge transfer, by measuring similarities across source and target distributions.

\section{Experimental Setup}\label{sec:expriment}

In this section, we describe the data and recommendation algorithm  that are at the base of our study. We describe a set of performance  metrics  that we use to evaluate user experience. We then discuss the construction behind our similarity measure that we use to transfer knowledge between different markets.

\subsection{Data}\label{sec:data}
We model the recommender system on real data with historical product purchases. We do experiments  on two dataset from different sources. Our primary dataset is made available to us from a 
retailer of  durable
consumer goods and it is collected from their online store. The second set is a publicly available dataset with purchase data.

A \textit{purchase} is as a sequence of item ID:s (representing products) bought by a user $\vect{x}=(x_1, x_2, x_3,\dots)$  with preserved  order of how the items where added to the shopping cart. We also refer to a purchase as a \textit{session}.
We remove purchases with a single item from the data, and cut sequences to contain at most 64 items.
Note that there is no user-profiling: the same user might generate several $\vect{x}$-sequences if the purchases are from different sessions, and there are no identifier  or information about the user attached to the sequences.

The online store 
operates in a number of  countries and the purchase data for our study is gathered  from twelve geographical markets.
The datasets have different sizes (total number of sequences $\vect{x}$) and the time-period for collection varies. There is also some slight variations in the product-range available in each market. Data from one market generally has 300,000--1,600,000 purchases  collected during a period no longer than two years, while between 10,000 and 20,000 unique items are available in each market. See also Table \ref{tab:freq} for markets used in experiments in Section \ref{sec:exp1}.

\begin{figure}
    \centering
    \includegraphics[width=\linewidth]{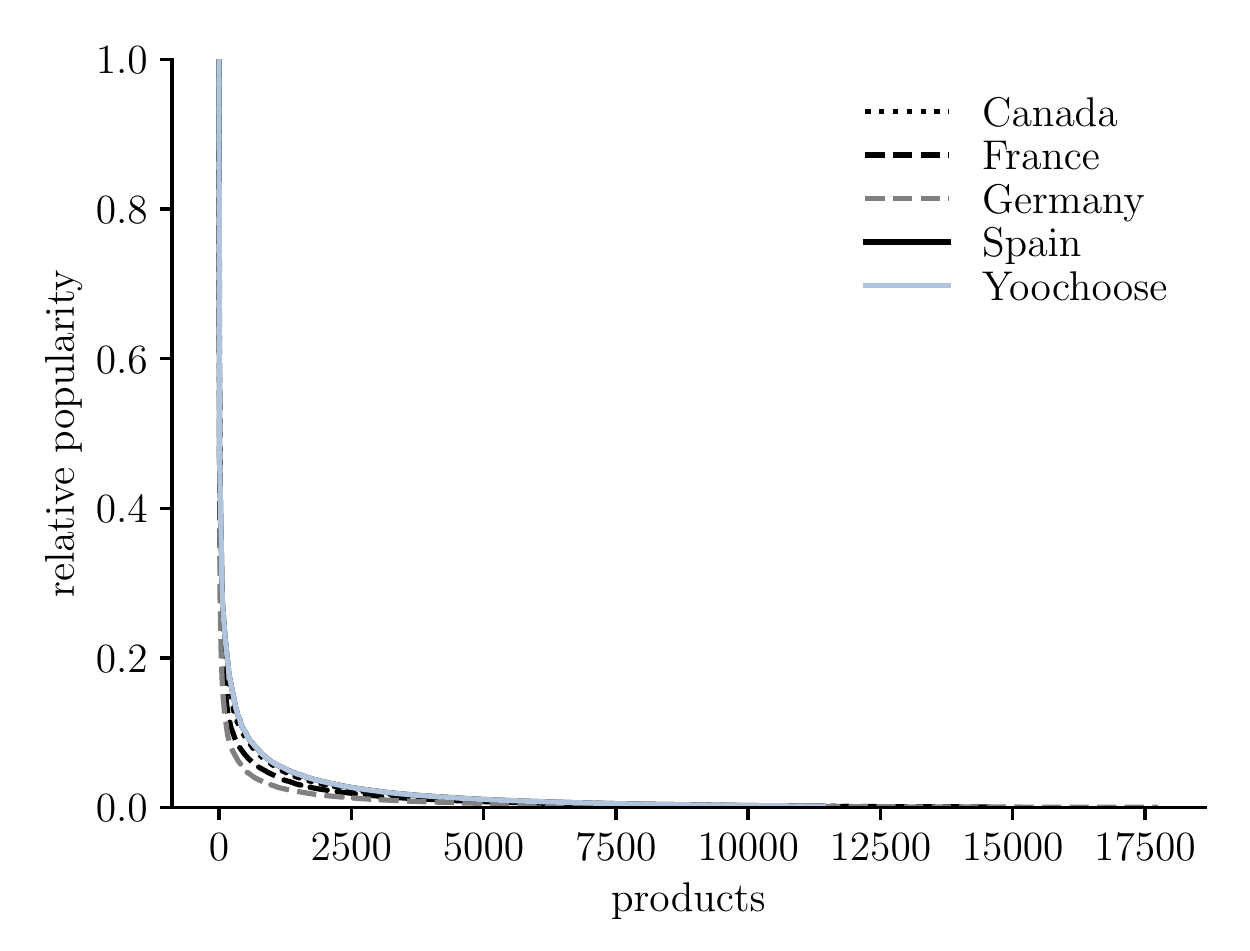}
    \caption{A comparison of the relative popularity distribution between the \textit{yoochoose} dataset and a selection of the datasets from the online store. The distribution of Span is hidden by the (solid blue) line of the \textit{yoochoose} distribution.  }
    \label{fig:distrubtion}
\end{figure}

As secondary data, we  use the publicly available \textit{yoochoose} dataset from the \textit{Recsys 2015} challenge \cite{recsys2015}. The purchase data in this dataset is comparable to a medium-sized market from the online store, such as Canada. It contains a similar amount of purchases ($510,000$ in \textit{yoochoose} and $490,000$ in the Canadian dataset), number of unique products ($14,000$ in \textit{yoochoose} and $12,000$ in the Canadian dataset), while it has a slightly less concentrated
\textit{popularity} distribution over how frequently a product is purchased, see Figure \ref{fig:distrubtion}. 

The popularity of an item %$x_i$, $p(x_i)$, 
in a particular market $M$ is defined by Equation \eqref{eq:pop} 
such that $p(x_i)=1$ for the most popular (most purchased) item in that market while $p(x_i)=0$ for item(s) not purchased at all:
\begin{equation}
    p(x_i)=\frac{ \sum_{x_j \in X_M}1_{\{x_j=x_i\}} }{\sum_{x_j \in X_M}1_{\{x_j=x_k\}}},\quad x_i\in X_\text{cat}.
    \label{eq:pop}
\end{equation}
Here $X_M$ is the dataset of all items {purchased in a market},
and $x_k$ the most popular product in that market. We use $X_\text{cat}$ to denote the product catalogue, i.e. the set of $n_x=|X_\text{cat}|$ unique items available in a market. 

\begin{figure}[ht!]
\begin{subfigure}{.45\textwidth}
\centering
\includegraphics[width=\linewidth]{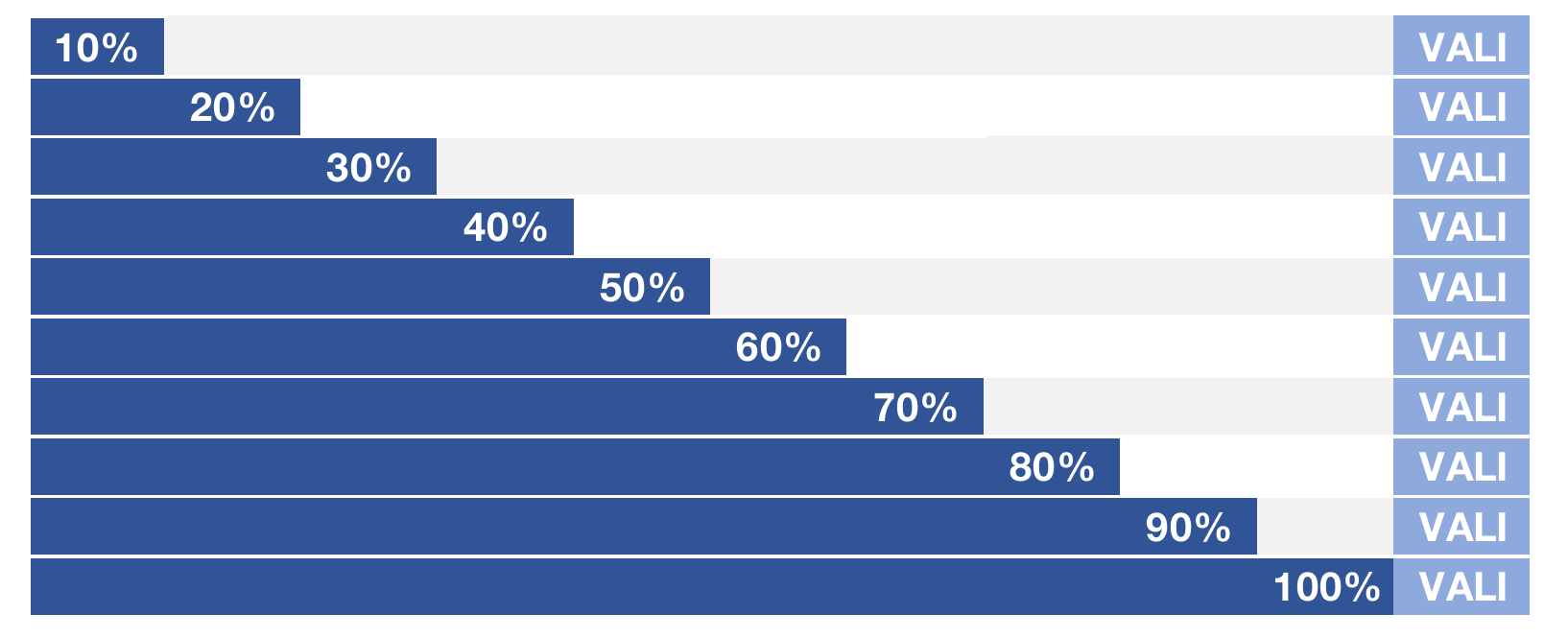}
\end{subfigure}
\caption{Partition of data. For each market, the validation set $X_\text{val}$ is $\nicefrac{1}{11}$ of the total data. The remaining $\nicefrac{10}{11}$ is the training data $X$. In the experiments, 10\%  to 100\% of $X$ is used for training on different sized datasets.   }
\label{fig:data_split}
\end{figure}

We partition the dataset of each market into a {validation} (\nicefrac{1}{11}) and training set (\nicefrac{10}{11}) by random assignment, i.e with no regard to chronology of when sessions where recorded (note that the set of sequences $x\in X_M$ is randomised, not individual items $x_i\in x$). This to avoid seasonality effects on our results. We further distribute the training set $X$ into 10 equally-sized buckets, and construct 10 new partitions by taking the first bucket as the first set (\nicefrac{1}{10} of $X$), the first and second bucket as the second set (\nicefrac{2}{10} of $X$) etc., see Figure \ref{fig:data_split}. 
This will be used in the experiments to asses how  validation performance depends on different sizes of  training data. Note that the validation set $X_\text{val}$ is kept the same for all partitions of the training set.

\subsection{Recommender algorithm}\label{sec:NN}
We use a recurrent neural network to model the sequential purchase data. This type of networks have been successfully applied to recommender systems and they are particularly well suited to session-based recommendations with no user profiling, \cite{hidasi2015session}. 

There are several variations and extensions of this model. As example, \cite{hidasi2016parallel} propose to augment the input space with feature information, \cite{tan2016improved} use data augmentation to improve performance while  a session-aware model is proposed in \cite{quadrana2017personalizing} with hierarchical recurrent network. 

\begin{figure}[h!]
    \centering
    \includegraphics[width=0.7\linewidth]{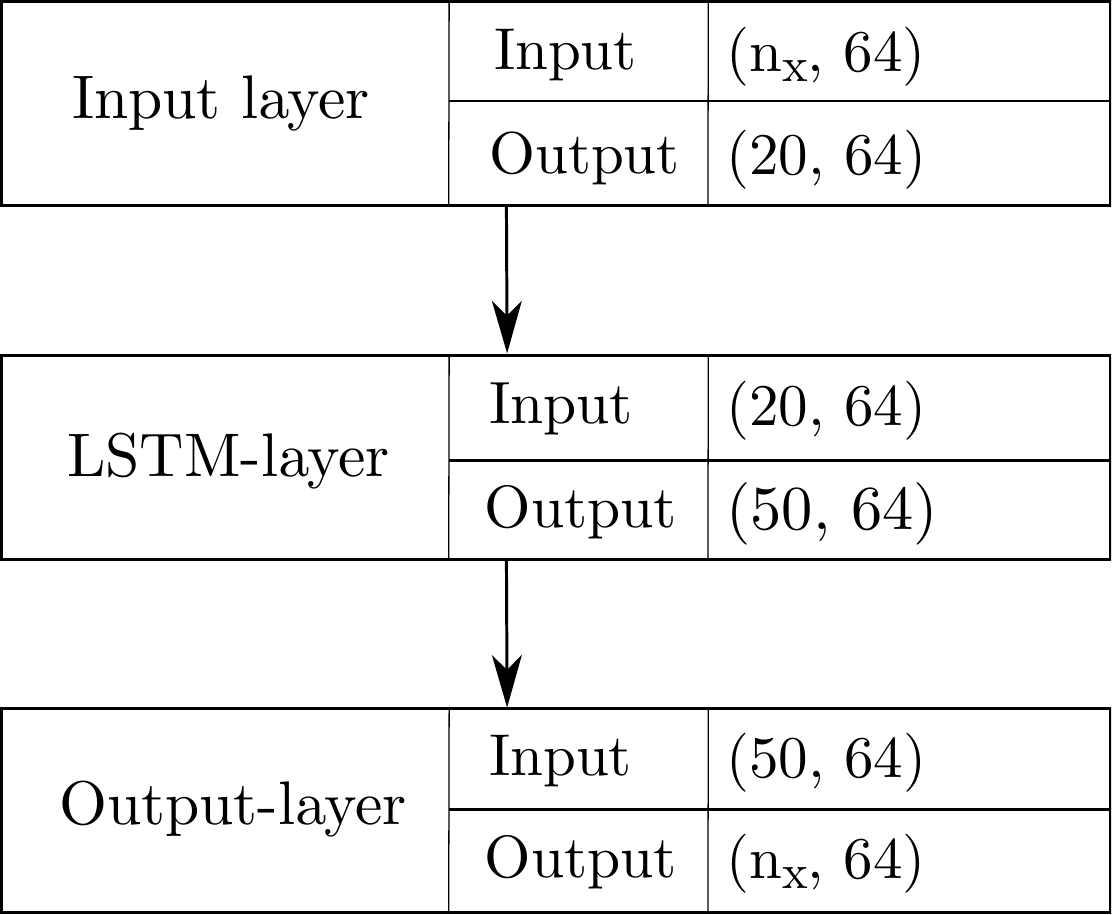}
    \caption{The network implementation with input and output dimensions of each layer. The input is a sequence of 64 items, each of which is a one-hot encoded vector of length $n_x$, the number of available items in the product catalog. We use a dimension of 20 for the embedding and 50 output units of the LSTM cell. The output of the network is a probability for each of the $n_x$ items. }
    \label{fig:lstm-impl}
\end{figure}

The high-level architecture of our network is shown in Figure \ref{fig:lstm-impl}. We use a long short-term memory (LSTM) model as the recurrent layer \cite{lstm}. A key mechanism in the LSTM is self-loops that creates stable connections between units through time, %over multiple scales, 
while the scale of integration is dynamically changed by the input sequence. This enable interactions over multiple time-scales and also prevents gradients from  vanishing  during training, see e.g. \cite{goodfellow2016deep}.

The (one-step) input to the network is an item  $x_t$ and the (one-step) output is a probability vector $\hat{y}_{t}$ for predicting the next item $y_t = x_{t+1}$ that will be added to cart. We use one-hot encoding for the items: $x_t$ (and $y_t$) is a column vector with length equal to the number of unique items $n_x$, where the element corresponding to the active item is 1 while the remaining $n_x-1$ elements are zero.

%  the training example is a sequence (explained below) but the one-step input is a one-hot encoded vector in my description (which is equivalent to just using integer encoded categories, and then a "look-up" operation)(this is completely equivalent to using Keras embedding layer, I just describe it in a formal and more general language - don't worry if it sounds too mathematical/technical, it should be on a good 'level' for this kind of publication)

Training takes a session $x$ as an example, where each item is used to predict  the subsequent item in the sequence. We use categorical cross entropy\footnote{The output vector $\hat{y}_t$ will have the same dimension $n_x\times 1$ as the one-hot encoded output target, and $\cdot$ denotes the dot product between two such vectors.   } 
\begin{equation}\label{eq:cce}
{l}(\hat{y}_{t}, {y}_{t}) = - y_t \cdot \log \hat{y}_{t}
\end{equation}
such that  $\mathcal{L}(\hat{y}, y) = \sum_{t=1}^{64} l(\hat{y}_{t}, {y}_{t})$ is the training loss of one example with 64 items. With $m$ training examples $x^{(1)},\dots,x^{(m)}$ collected from a market, the total cost
\begin{equation}\label{eq:cost}
J_m(\theta)=\sum_{i=1}^m \mathcal{L}(\hat{y}^{(i)}, y^{(i)})
\end{equation}
is used for optimisation objective during training of the parameters $\theta$ in the network.

The first hidden layer of the network is an embedding that maps each item to a real valued $n^{[1]}$-dimensional column vector $$h_t^{[1]}=W^{[1]}x_t$$ where $W^{[1]}$ is an $n^{[1]}\times n_x$  weight matrix. If the $j$:th element in $x_t$ is active,  this is a simple look-up operation where $h_t^{[1]}$ is a copy of the $j$:th column of $W^{[1]}$. 
The layer thus outputs a continuous vector representation of each categorical input: it \textit{embeds} each (on-hot encoded) item ID into the geometrical space $\mathbb{R}^{n^{[1]}}$. Importantly, the embedding is learned from data such that distances in $\mathbb{R}^{n^{[1]}}$ have a meaning in the context of our learning problem. Items that are commonly co-purchased will be close to each other in the embedding space. We will further exploit this property %to define similarity measures between markets 
in Section \ref{sec:sim}.

The second hidden layer is a LSTM cell that maps the current input and its  previous output (this is the recurrent feature) to an updated output: $$\left( h_t^{[1]}, h_{t-1}^{[2]}\right) \mapsto h_{t}^{[2]}$$ where $h_t^{[2]}$ is an $n^{[2]}$-dimensional vector. The key component is an internal state $s$ that integrates information over time with a self-loop. A set of gates $f$, $g$ and $q$ respectively control the flow of the internal state, external inputs and new output of the layer:
\begin{align*}
    s_t &= f_t \circ s_{t-1} + g_t \circ \sigma\left( Uh_t^{[1]} + Vh_{t-1}^{[2]} + b \right), \\
    h^{[2]}_t &= q_t \circ \tanh\left( s_t \right).
\end{align*}
$U$, $V$ and $b$ are input weights, recurrent weights and biases; $\sigma$ is the sigmoid activation and $\circ$ denotes an element-wise product. All gates have the same construction. For the forget gate
\begin{equation*}
    f_t = \sigma\left( U^fh_t^{[1]} + V^fh_{t-1}^{[2]} + b^f \right)
\end{equation*}
where superscript indicates that weights and biases are associated with $f$. A corresponding $(U^g, V^g, b^g)$ and $(U^q, V^q, b^q)$ are used for the input and output gate respectively. Due to the sigmoid, the gates can take a value between 0 and 1 to continuously turn off/on  the interaction of their variable, and the gating is also controlled by the inputs to the layer. We use $W^{[2]} = (U,V,U^f,V^f,U^g,V^g,U^q,V^q)$ to collect all weight matrices of the layer, and $b^{[2]} = (b,b^f,b^g,b^q)$ for biases. These are  $n^{[2]}\times 1$, while $U$-matrices are $n^{[2]}\times n^{[1]}$ and $V$-matrices  $n^{[2]}\times n^{[2]}$ since they operate on the recurrent state.

The last hidden layer is a densely connected layer with a softmax activation $\gamma$ to output a probability vector:
\begin{equation}\label{eq:output_layer}
    \hat{y}_t = \gamma\left( W^{[3]}h_t^{[2]} + b^{[3]} \right).
\end{equation}
This is a probability distribution over all $n_x$ available products: $\hat{y}_t$ has the dimension $n_x\times 1$ (the same as the target $y_t$),
and its $j$:th element is the probability that the next item is the $j$:th product according to the one-hot encoding.  

The parameters in the network are collected with $$\theta = (W^{[1]},W^{[2]},b^{[2]},W^{[3]},b^{[3]}).$$
We learn $\theta$ from a training set of sessions $X=\{x^{(1)},\dots,x^{(m)}\}$ by minimising the total cost \eqref{eq:cost} on $X$ with the  Adam optimiser \cite{kingma2014adam}. 

Recommendations by the trained network are based on the next-step prediction $\hat{y}_t$ from a user's session $(\dots,x_{t-2}, x_{t-1}, x_t)$ up to the most recent item $x_t$ that was added to cart. For recommending a single item, we use the maximum probability in $\hat{y}_t$. Similarly, for a ranked list of top-$k$ recommendations, we take the items that have the $k$ largest probabilities. We denote this list with
$$\text{rec}_k(\hat{y}_t)\equiv \text{rec}_k(x_{t+1})$$
and use $\text{rec}_k(x_{t+1})$ when we want to emphasise that the target for the recommendation  is the (true) item $x_{t+1}$.

\paragraph{Learning-to-rank loss} While  categorical cross entropy \eqref{eq:cce} is the default loss for multiclass classification with neural networks, it is not the only choice. For recommender systems, there are several alternatives and \textit{learning to rank} criteria are popular since the recommender's problem is ultimately ranking a list of items. A challenge with  cross entropy for recommendations is also the sparse nature of the classification problem. Processing a training example $(...,x_{t-1},x_t, y_t)$ will update  parameters associated with only one node of $\hat{y}_t$, the prediction probability for the active, `positive' item of $y_t$. Predictions of all the other $n_x-1$  `negative' items will not be improved by the training example. Moreover, since the distribution of how often items are active is typically very skewed in data (see Figure \ref{fig:distrubtion}), the model is mostly  trained at predicting (the small set of) popular items with many positive observations. 

To this end, Bayesian personalised ranking (BPR) is a successful criterion for recommendations that optimise predictions also for negative items \cite{rendle2012bpr}. It considers pairwise preferences for a positive and a negative item, such that the model score is maximised for the former and minimized for the latter. In the experiments, as complement to cross entropy, we use a BPR loss adapted to recurrent neural networks %as proposed 
by \cite{hidasi2018recurrent}
\begin{equation}\label{eq:bpr_loss}
{l}_\text{BPR}(\hat{y}_{t}, {y}_{t}) = -\log \sum_{j \in \mathcal{N}_S} \hat{y}_{t}[j]\sigma\left( z_t[i] - z_t[j]\right). 
\end{equation}
$\mathcal{N}_S$ is a set of uniformly sampled indices of negative items and $\hat{y}_t[j]$ is used to denote the $j$:th element of $\hat{y}_t$ while $i$ is the positive element, i.e. the active element of $y_t$. The model score $z_t = W^{[3]}h_t^{[2]} + b^{[3]}$ is the pre-activation of the output layer \eqref{eq:output_layer} and $\sigma$ is the logistic sigmoid.

\paragraph{Graph based recommender algorithm}

% We also tested a graph based recommender algorithm where the product catalog was modeled as a Markov chain. Recommendations was then generated from this markov chain using a random walk algorithm. 

For comparison, we also use a graph-based  algorithm where purchase sessions are modeled by a Markov chain with discrete state space. %Recommendations are generated by a random walk algorithm. 
Each unique item in the product catalog is a state in the Markov chain, and the transition probability between a pair of products is estimated by maximum likelihood, i.e. from how frequently the two products occur consecutively in a purchase of the training dataset. %, $X_\text{train}$.

% \begin{equation}
%         {P}(x_k,x_m)=\frac{\sum_{\textup{p}_i \in X_\text{train}} 1_{\{x_k\in\textup{p}_{i},x_m\in\textup{p}_{i} \}}}{\sum_{x_j \in X_\text{cat}}\sum_{\textup{p}_i \in X_\text{train}} 1_{\{x_k\in\textup{p}_{i},x_j\in\textup{p}_{i} \}}} 
% \end{equation}

% To generate recommendations based on items that has been added to the shopping cart, $X_\text{cart}=\left \{ x_1,x_2..., x_n \right \}$ we use a random walk on the markov chain. The random walk starts at each of $x_k \in X_\text{cart}$ and taking 2 steps in the product graph (two transitions). This is repeated 500 times. 

% We select the $k$ products that have occurred most frequently in all random walks from all products as recommendations for the user. 

To generate a list of $k$ recommendations based on items that has been added to the shopping cart, $X_\text{cart}=\left \{ x_1,x_2..., x_n \right \}$, we use a random walk according to the Markov chain. We start the walk at each item in cart $x_k \in X_\text{cart}$ and take two random steps  (two transitions). This is then repeated 500 times. We select the $k$ products that have occurred most frequently in all random walks from all products as recommendations for the user. 

\subsection{Performance metrics}
%--
Performance of multiclass classification algorithms, such as  recommender systems, is commonly evaluated with metrics of accuracy, recall or F-measures. However, these do not necessarily give a good representation of how satisfied the user is with item suggestions, see e.g. \cite{10.1145/2487575.2488215, 10.1145/2645710.2645745}.  Therefore, to better quantify performance in terms of user experience, we use a combination of three metrics for evaluation: top-K accuracy, catalog coverage and novelty. %is important but insufficient to quantify how satisfying a user experience is. To better quantify the quality of the user experience we use three different metrics in this paper: Top-K accuracy, Catalog Coverage and Novelty.

Top-K accuracy %, as defined in  \eqref{prek}, %and \ref{recdef}, is related to accuracy in the sense that it 
is a proxy measure of how often suggestions by the recommender system aligns with the actual item selected by the user, see e.g. \cite{kim_book}. The algorithm predicts a list of $k$ items for the next item of the user's purchase. If any of these is the true next item, the recommendation is considered successful. Top-K accuracy is  then the ratio of successes on a validation dataset $X_\text{val}$
\begin{equation*}
    \text {top-K accuracy}=\frac{1}{|X_\text{val}|} \sum_{x_i \in X_\text{val}} 1_{\{x_i\in\textup{rec}_{k}(x_i)\}},
    \label{prek}
\end{equation*}
% \begin{equation*}
%     \text{hit}_k(x_i) =\left\{\begin{matrix}
%  1, & \text{if the $k$ recommendations include the true item $x_i$}   \\ 
%  0, & \text{otherwise},
% \end{matrix}\right.
% \end{equation*}
where $1_{\{x_i\in\textup{rec}_{k}(x_i)\}}$ is one if the $k$ recommendations include the true item $x_i$, zero otherwise, and $|X_\text{val}|$ is the cardinality of the validation set. 
%As described in the previous section, previous  items  up to $x_t$ are used for predicting a list of $k$ items for $x_i=x_{t+1}$.
%where $L$ is the list of recommendations on a validation set. Note that each recommendation $i$ contains $k$ items from $L$. 

Catalog coverage, as defined in Equation \eqref{cc}, 
is the proportion of items in the product catalog $X_\text{cat}$ that are actively being suggested by the recommender algorithm on a validation set, see e.g. \cite{beyondacc}. A high coverage provides the user with a more detailed and nuanced picture of the product catalog. 
\begin{equation}
    \text { catalog coverage }=\frac{1}{|X_\text{cat}|} \left|\left\{ \cup_{x_i\in X_\text{val}} \text{rec}_k(x_i) \right\}_{\neq} \right| %\left|\cup_{i=1, |L|} L_i\right|}
    \label{cc}
\end{equation}
where $\{\cdot\}_\neq$ is used to denote that the cardinality is taken over the set of all {unique} items in the union of recommendations.

Novelty is a metric that aspires to capture how different a recommendation is compared to items previously seen by the user;  %what the user has previously seen
\cite{nov_and_div}. Since our data %analyzed in this work 
does not contain user identifiers, it is no possible to say which items have been viewed by the user before the %a particular 
purchase. %if an item has been viewed by the user before. 
Instead, we use the popularity of an item \eqref{eq:pop} as a proxy and define the metric as  %as a measurement for novelty. Less popular items are therefore premiered in this metric. Novelty is defined in Equation \eqref{novelty} where $p(x_i)$ is the relative popularity (between $0$ and $1$) of an item $x_i$ in the dataset. {\color{yellow}ref to figure 1}
\begin{equation}
    \text{novelty} =-\frac{1}{|X_\text{val}|\cdot k}\sum_{x_i \in X_\text{val}} \sum_{l=1}^k \log p\left(\text{rec}_k(x_i)_l \right)
    \label{novelty}
\end{equation}
where $p(\text{rec}_k(x_i)_l)$ is the relative popularity (between $0$ and $1$) of the $l$:th item in $\text{rec}_k(x_i)$, the list of $k$ items recommended for $x_i$. Less popular items are therefore premiered in this metric.

\subsection{Similarity measure}\label{sec:sim}
%--
As for the online store, it is common that a retailer operates in several related markets. In this case, 
%When the recommender has access to data from  domains other than the target domain (from which she collects data for development of the system), 
there is good reason to use knowledge transfer to leverage  data across markets.  We propose a similarity measure for this purpose, to better   judge which `foreign' market data (source domain) is suitable for developing a recommendation system in the `domestic' market (target domain).

One possible approach to measure data compatibility between %markets
domains is by using %deep learning 
embeddings. %Embeddings are mappings between variables, in this case purchases, to numerical vectors. 
These are continuous representations of the data that are  inferred  from the  learning problem. In particular, the learned representation has \textit{meaning} such that geometry in the embedding space 
corresponds to contextual relationships. A striking example is the embedding technique of the word2vec model, which successfully captures semantic similarities by distances in the embedding space  \cite{mikolov2013efficient}.
% Techniques using embeddings has proved to be very successful in capturing an item's contextual meaning. An example of this is the famous NLP technique Word2Vec which is successfully able to describe semantics of words. 
%We attempt to develop 

We use a model for the recommender system with an embedding layer that captures purchase behaviour in a  markets (the first hidden layer in the neural network, see Section \ref{sec:NN} ). Distances in %among 
these embeddings of different markets are at the base of our similarity measure.

%We utilize the LSTM model used for the recommender system, as pictured in figure \ref{fig:lstm-impl}, to obtain these embeddings. 
To construct the measure, we first train the network on a dataset with data from all markets, to obtain a global representation with no regional dependencies. %When the training of the model was completed, 
We then remove all layers except for the initial embedding layer from the model. The input to this layer is an (encoded) item ID and the output is a vector with length $n^{[1]}=20$ that represents the item in the embedding space. %, a embedding, for each of the $64$ input items.
For a purchase  $x$ with 64 items, we concatenate the corresponding 64 embedding vectors %of length $20$ 
to obtain a vector $\mathbf{h}$ of length $1280$. We take this as a (global) representation of a particular purchase. 
%--

We use these vectors to %compute a general representation of 
measure similarities between  purchase behaviour in  different markets. %This is accomplished by taking
We take $850$ purchase vectors from each market, %that we have interest in and then computing their corresponding embeddings. From these 850 embeddings per market we
and compute a centroid from Equation \eqref{cent}. %, see \cite{centroid}. %, using the definition in equation \ref{cent}, and 
We let this centroid represent the %general 
(local) purchase behaviour in that market. 

$\mathbf{\bar{p}}_M$ %in equation \ref{cent} 
is the computed centroid for market $M$, $\mathbf{h}^{(i)}$ is the representation for purchase $x^{(i)}$ in that market $M$, and $n=850$ is the number of such embedding vectors used to compute the centroid:
\begin{equation}
    \mathbf{\bar{p}}_M=\frac{\sum_{i=1}^n \mathbf{h}^{(i)}}{n}
    \label{cent}
\end{equation}
Finally, we compute the similarity between two %the embedding representation of 
markets with $\mathbf{\bar{p}}_{M_1}$ and $\mathbf{\bar{p}}_{M_2}$ by  cosine similarity: %, as detailed in equation \ref{cosinesim}.
\begin{equation*}
    \textup{similarity}=\frac{\mathbf{\bar{p}}_{M_1}\cdot\mathbf{\bar{p}}_{M_2}}{\left \|\mathbf{\bar{p}}_{M_1}  \right \|\cdot \left \|\mathbf{\bar{p}}_{M_2}  \right \|}.
    \label{cosinesim}
\end{equation*}

\begin{figure}[ht!]
\begin{subfigure}{.42\textwidth}
  \centering
  % include first image
  \includegraphics[width=\linewidth]{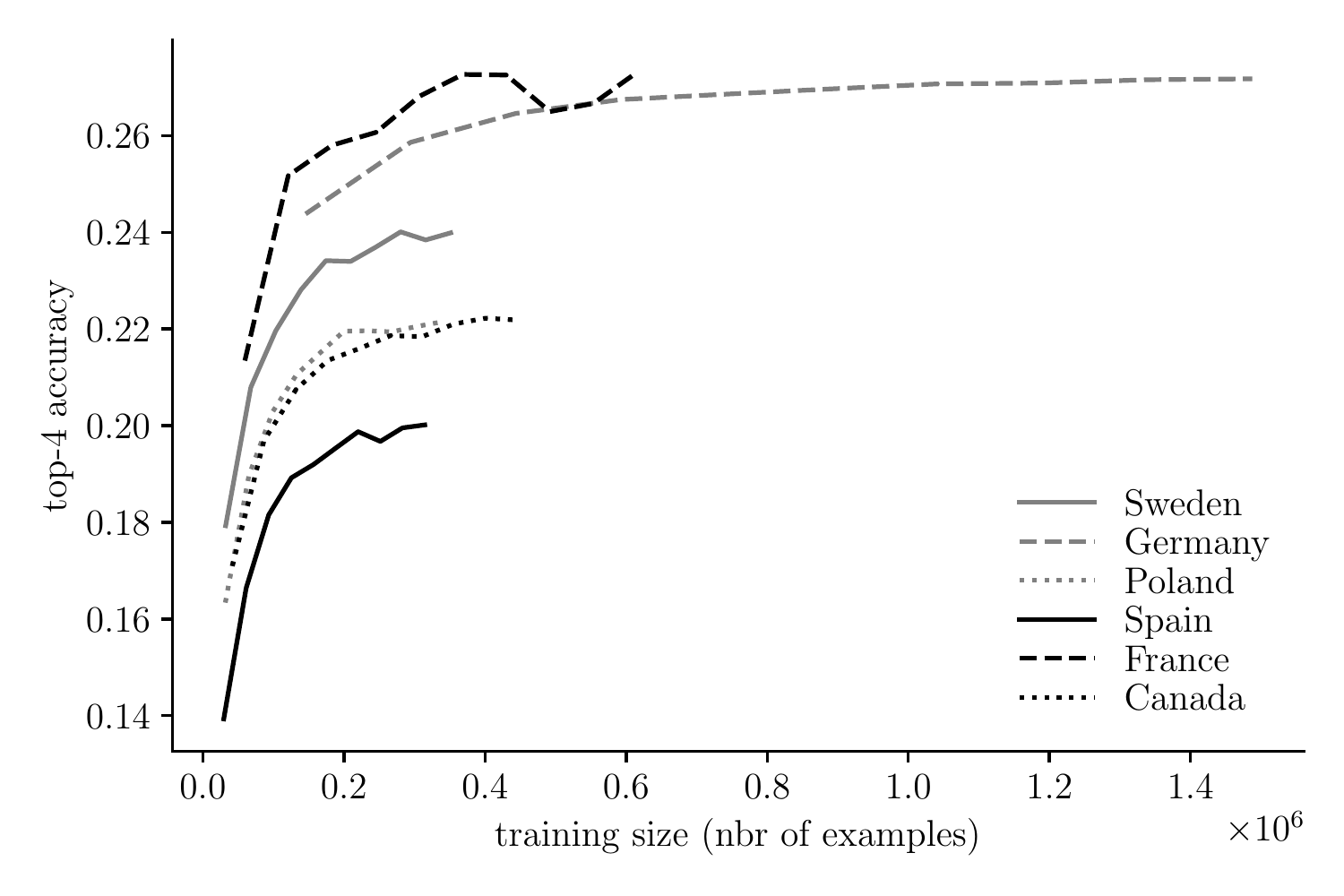}  
  \caption{}
  \label{img:1pre@k}
\end{subfigure}
\begin{subfigure}{.42\textwidth}
  \centering
  % include second image
  \includegraphics[width=\linewidth]{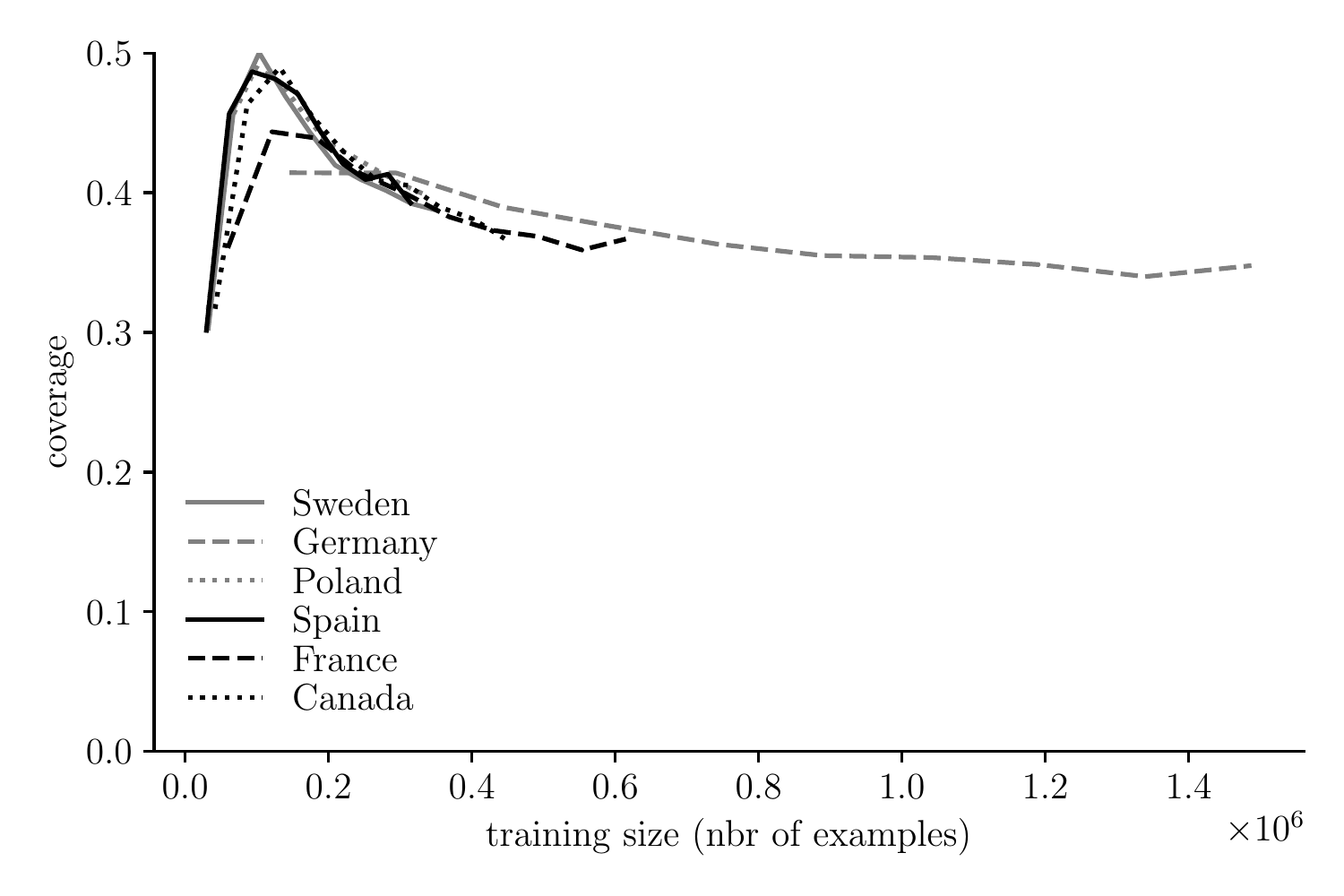}
  \caption{}
  \label{img:1cov}
\end{subfigure}
\begin{subfigure}{.42\textwidth}
  \centering
  % include second image
  \includegraphics[width=\linewidth]{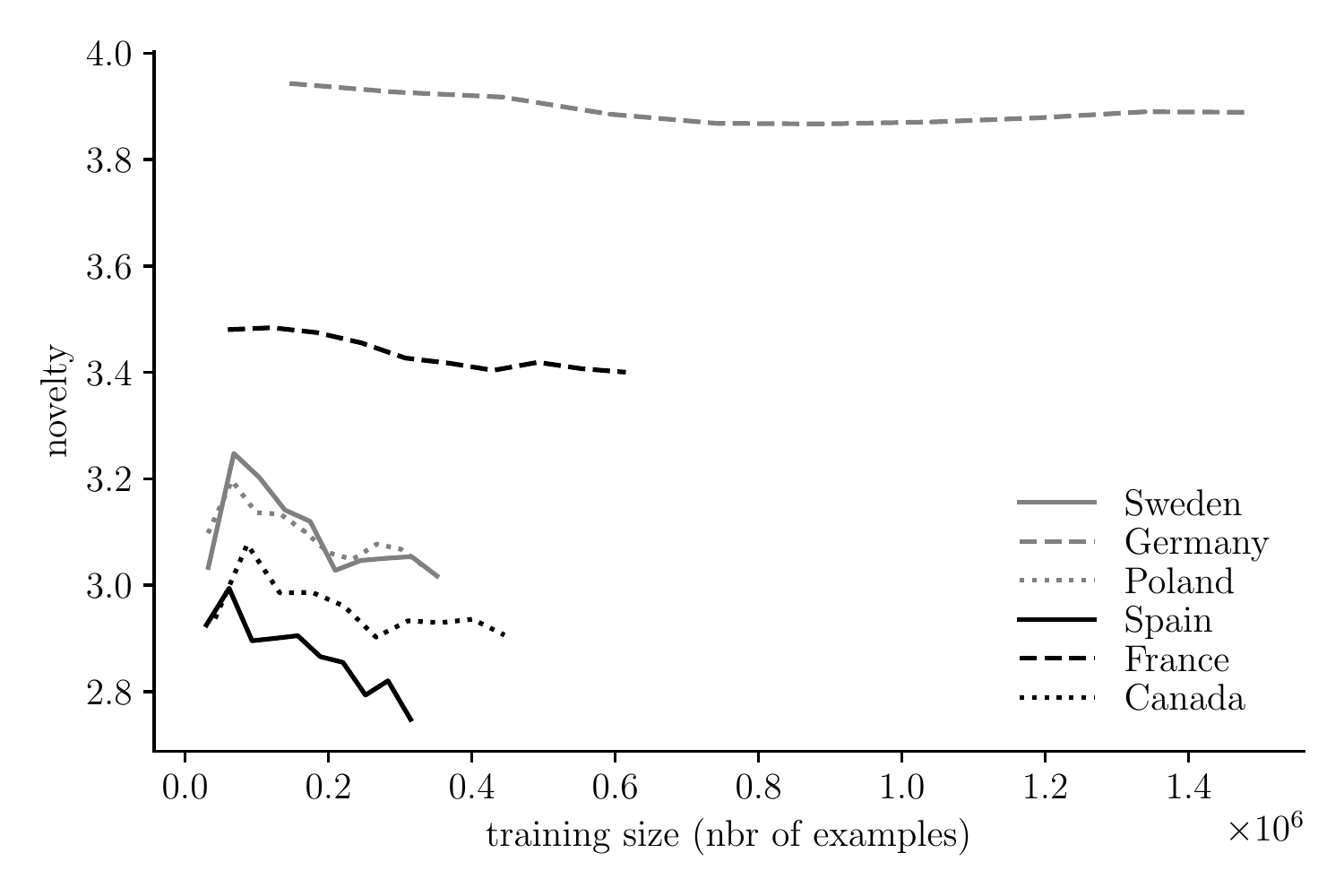}
  \caption{}
  \label{img:1nov}
\end{subfigure}
\begin{subfigure}{.42\textwidth}
    \centering
    \includegraphics[width=\textwidth]{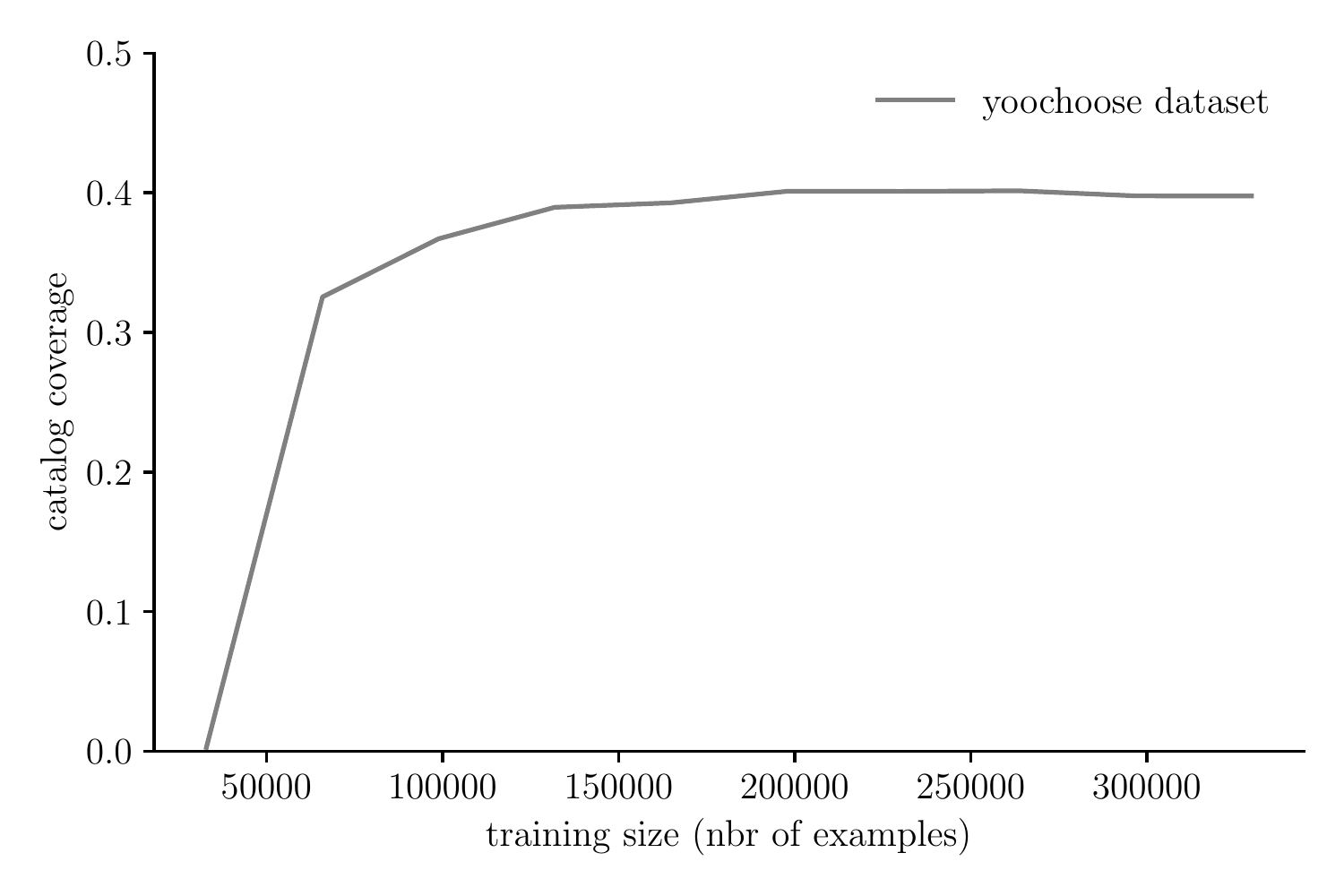}
    \caption{}
    %Catalog coverage validated on the \textit{yoochoose}-dataset. 
    \label{fig:yoochoose-cov}
\end{subfigure}
\caption{Validation performance in top-4 accuracy (a) catalog coverage (b) and novelty (c)  as a function of training data size %. Figures (a)-(c) show results 
from six different markets of the  online store. Figure (d) shows catalog coverage validated on the \textit{yoochoose}-dataset. }
\label{img:res1}
\end{figure}

\section{Experimental Results}\label{sec:expRes}
We conduct two sets of experiments in the section. In the first, we investigate how the amount of training data affects the recommendation system's performance. In the second, we investigate if training data from secondary sources can improve performance, and if our similarity measure can be useful in the selection of such data.

\subsection{Performance of size of training data}\label{sec:exp1}
In this experiment, we analyse how performance on the validation data varies as we include more and more examples in the training set used for learning  network parameters. The training data is partitioned as explained in Section \ref{sec:data}. %; see also Figure \ref{fig:data_split}. 
Note that we use the same hold-out validation set for each training set, it is just the amount of training data that is changed (increased in size). For the network, we use the architecture in Figure \ref{fig:lstm-impl} with an embedding dimension $n^{[1]}=20$ and $n^{[2]}=50$ hidden units of the LSTM cell.  In the learning process, we also keep a constant setting for each training set: we optimise with mini-batch gradient descent, with batch size of 64 examples  and run for 25 epochs. When we increase the size of the training set, we cold-start the optimisation with uniform Xavier initiation.
%We train 10 models on different sized data sets and evaluate using our specified metrics. We evaluate these 10 models trained on different sized training data sets using the same validation set. 

We repeat the experiment on data from the online store on six different market to further nuance the results (with six separate networks) as well as on the yoochoose dataset; see Table \ref{tab:freq}.
\begin{table}
\small
  \caption{Description of data.}
  \label{tab:freq}
  \begin{tabular}{ccc}
    %\toprule
    \hline
    Market &  Data (\#purchases) & Catalog (\#items) \\
    % \midrule
    \hline
    Canada & 490,000 & 12,000 \\
    France & 676,000 & 15,0000 \\
    Germany & 1,636,000 & 18,000 \\
    Spain & 346,000 & 11,000\\
    Sweden & 386,000 & 11,000\\
    Poland & 368,000 & 10,000\\
    Yoochoose & 510,000 & 14,000 \\
%   \bottomrule
    \hline
\end{tabular}
\end{table}
We report on all three  metrics to evaluate  performance, and use a list of $k=4$ items for top-K accuarcy. 
%To split the available data from one market into a validation set and 10 training sets, first, all available data for that specific market was extracted and divided randomly into 11 sets. One of the sets was selected to be the validation set for that market. Then the 10 remaining sets were combined to form the 10 training sets of different sizes. 
%--
Results are shown in Figure \ref{img:1pre@k}-\ref{img:1nov} for the primary data, and in Figure \ref{fig:yoochoose-cov} for the secondary data (we only include a plot of the catalog coverage metric).  

From figure \ref{img:1pre@k} we  observe a clear  tendency that as the amount of training data increases, top-4 accuracy on validation data (quickly) increases up to a threshold before it levels off. This  logarithmic behavior of the increase suggests that there is a law of diminishing returns due to the irreducible generalisation error that cannot be surpassed with larger amounts of training data \cite{baidu}. We see this for all studied markets. This is on a par with the saturation effect reported in \cite{larson2017towards}. However, while they conclude a decline of accuracy with the  squared error metric on training data, we look at {validation performance}, with metrics purposely designed for measuring quality of recommendations.

The overall difference in performance between  markets  is most likely due to how much the purchasing patterns \textit{vary} within a market; i.e. the degree of variability---\textit{entropy}---in the generating data distribution. If purchasing patterns vary more within a market, predicting them will be less trivial. Accuracy can also be connected to the  popularity distribution of a market, shown in Figure \ref{fig:distrubtion}. For example Germany achieves high levels of top-4 accuracy around 0.27  while its popularity distribution is concentrated over relatively few products. The network can then achieve high accuracy by concentrating its recommendations on a smaller active set of popular items. Similarly, Spain has a `flatter' popularity distribution as compared to other markets, such that the network has to learn to recommend a larger set. It has an top-4 accuracy that levels off around 0.2.

For the \textit{yoochoose} data, a similar decline in increasing validation performance is observed while the network reaches a higher level of top-4 accuracy around 0.43. This indicates that purchases are much easier  to predict in general in this dataset: since its popularity distribution is very similar to Spain (see Figure \ref{fig:distrubtion}) it has a relatively large active set of popular items. Still, it learns to recommend this set with  high overall top-4 accuracy.  

% \begin{figure}
%     \centering
%     \includegraphics[width=0.45\textwidth]{img/publi_result_cov.pdf}
%     \caption{Catalog coverage validated on the \textit{yoochoose}-dataset. }
%     \label{fig:yoochoose-cov}
% \end{figure}

Figure \ref{img:1cov} shows an opposite relationship between catalog coverage and the amount of training data. Catalog coverage on the validation set is  low for the smallest sized training sets. When reaching a sufficient amount of data, the network learns to `cover' the catalog, and the metric peaks. After this peak, catalog coverage decreases as more training data is added. This is observed for all markets.
However, we observe a different pattern on the \textit{yoochoose} dataset, see Figure \ref{fig:yoochoose-cov}. On this dataset we  see a law of diminishing returns, similar to what we observed for top-K accuracy in Figure \ref{img:1pre@k}. An explanation for this  behaviour could be that the data distribution has less variability, such that the \textit{yoochoose} dataset is more predictable, which the high level of top-K accuracy indicate. If the dataset is easier to predict, the recommender system could more accurately predict items from the active set of items while the network is regular enough to cover the full product catalog. In contrast, since the datasets from the online store seem to be less predictable, the recommender system learns to recommend the most popular items when it is trained to more data. The network is less regular such that it leaves the less popular items out, which thereby lowers catalog coverage.  

Again, validation performance in terms of catalog coverage makes a good case for preferring  minimal necessary data: If catalog coverage is an important metric for the recommender, there is indeed a trade-off and even \textit{decrease} in performance  when the amount of the training data is increased.

Figure \ref{img:1nov} shows novelty on the validation set as a function of the amount of training data. The  impact on novelty from training size is less clear than for top-K accuracy and catalog coverage: For all markets except Germany, novelty  decreases when the amount of training data increases. On the \textit{yoochoose} data, novelty is rather constant with no clear effect from the size of the training set. Recommendations are generally more novel in the French and German markets. These are also the two markets with more concentrated popularity distributions, such that less popularity is assigned to a larger portion of the recommended items in the metric. This is probably a contributing factor to their high  levels of validation novelty.

Performance in terms of novelty  is quite robust for a couple of markets and \textit{yoochoose}, while we see a general decline in nolvelty on the validations set for the other markets. This indicates yet again that there is a (positive) trade-off between performance and training data size. 

In all, we see a strong case and good reasons for a view towards minimal %and good reasons for an analysis of 
necessary training data when developing and deploying a recommender system. This in all considered metrics, with the \textit{effect} of training size intervened with the underlying data distribution and typical purchase behaviour of users. For top-K accuracy, there is a \textit{saturation} and thus diminishing marginal performance when we increase the amount of training data. For catalog coverage there is a \textit{decline} in performance for the market data from the online store, and saturation on the more predictable \textit{yoochoose} data. Similarly, we also see declining performance in novelty for markets with a larger active set of popular item, while the metric is relatively constant in the other markets and the \textit{yoochoose} data.

%the largest datasets of available training data. %For all markets except Germany, novelty decreases when the amount of training data increases. In the German market, where the novelty overall is notably higher than in other markets. The difference in novelty is too small to draw any conclusions about how it is affected by the amount of training data. 

\begin{figure}[ht!]
\begin{subfigure}{.42\textwidth}
  \centering
  \includegraphics[width=\linewidth]{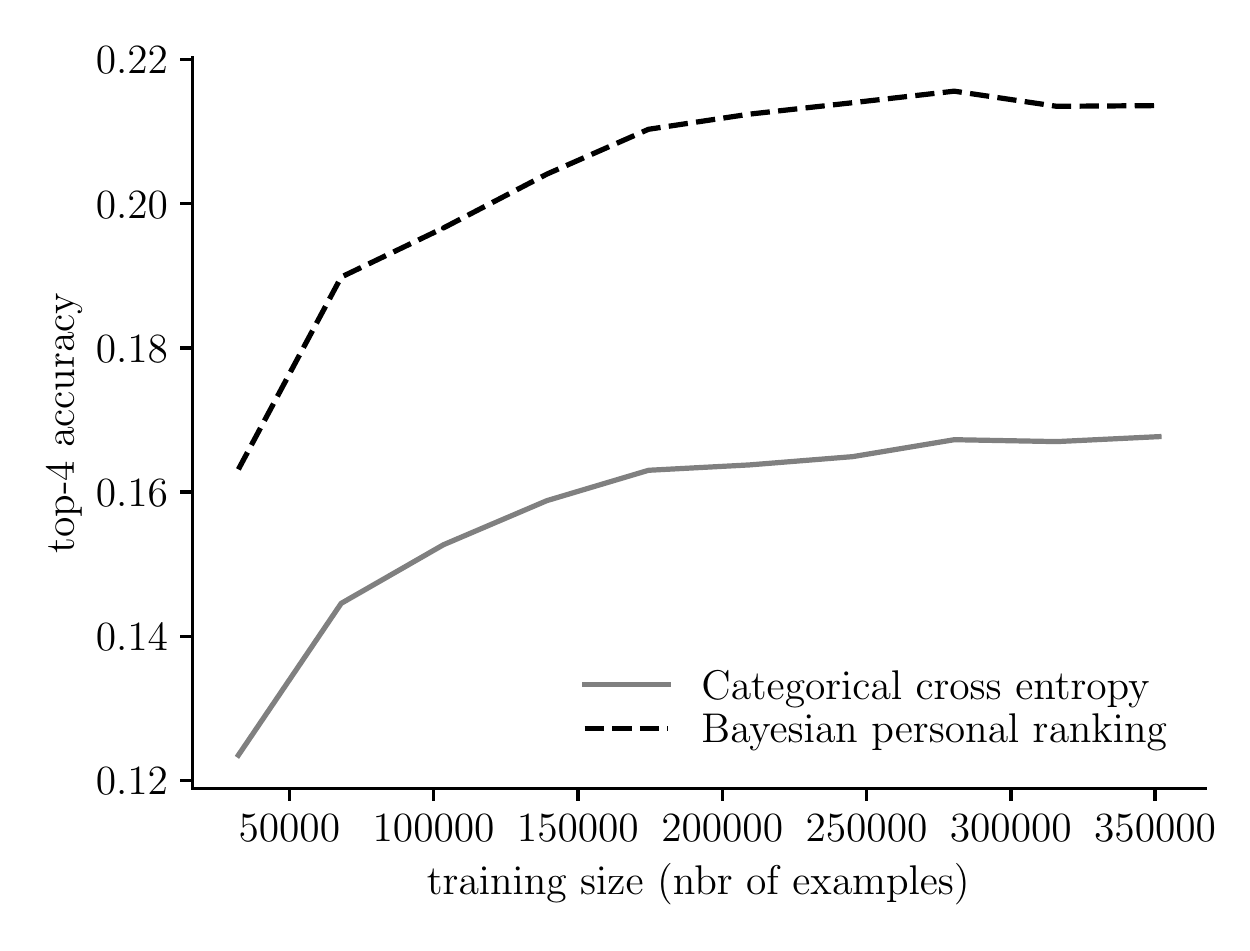}  
  \label{img:bpr_acc}  
  \end{subfigure}
  \begin{subfigure}{.42\textwidth}
\includegraphics[width=\linewidth]{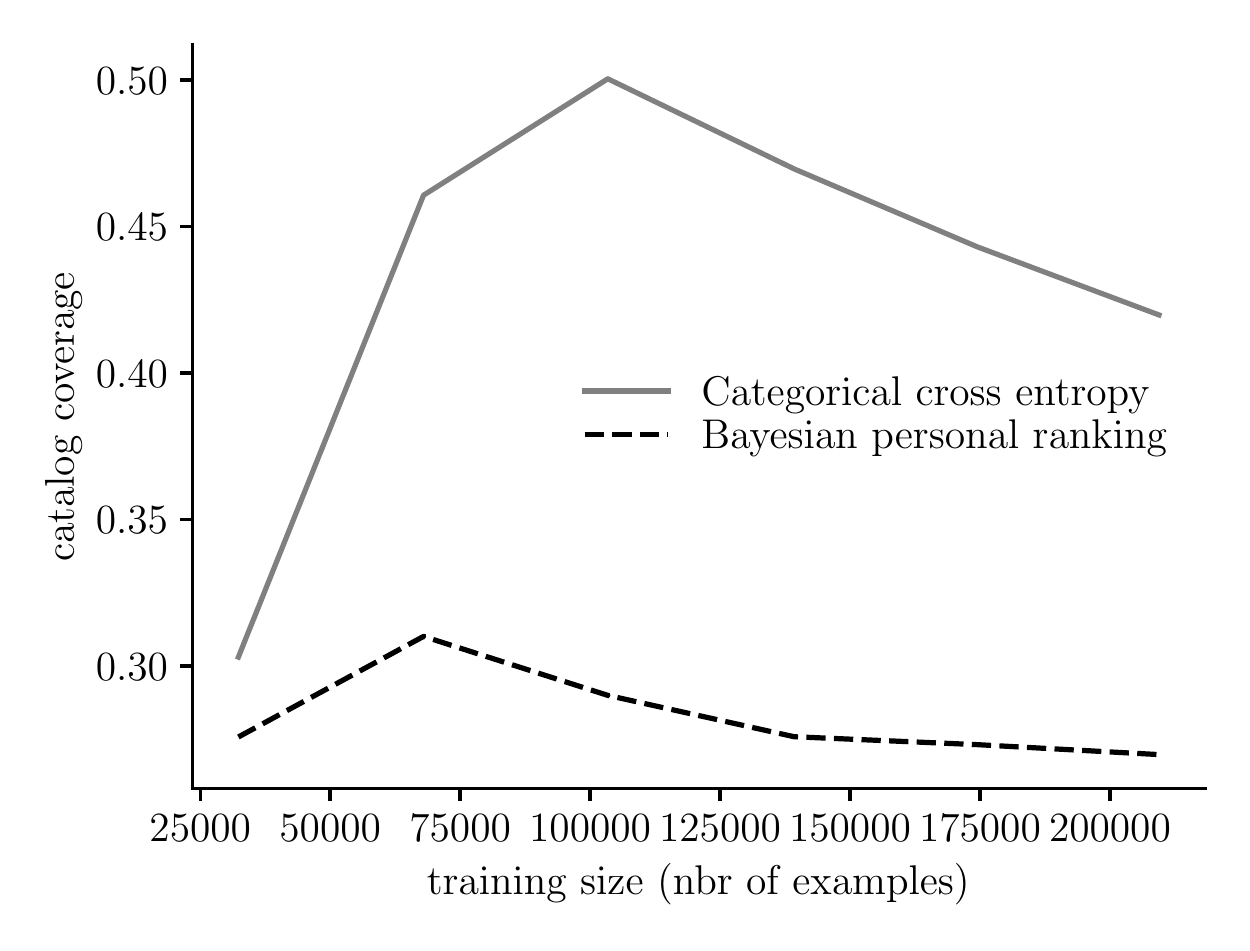}  
  \label{img:bpr_cc}
\end{subfigure}

\caption{Validation performance when training with Bayesian personalised ranking loss on Swedish data: top-4 accuracy (left) and catalog coverage (right).  }
\label{img:res_bpr}
\end{figure}

\paragraph{Ranking loss.} To complement the analysis, we repeat the experiment and use Bayesian personalised ranking loss \eqref{eq:bpr_loss} for training to the Swedish market. %instead of cross entropy. 
We keep the same model parameters and architecture (except for the softmax activation of the output layer), and use a sampling size of $|\mathcal{N}_S| = 1024$ negative items. For the resulting validation performance show in Figure \ref{img:res_bpr}, the trends from the above analysis continuous to hold. %Figure \ref{img:res_bpr} show validation performance. 
As for cross entropy, there is a clear saturation in top-4 accuracy for BPR. As promised by the ranking criterion, the overall accuracy is higher at the same amount of data, but notably, the \textit{return} on additional data is the same for the two losses: there is positive shift in accuracy when training with BPR. In return, there is a negative shift in catalog coverage. We still see a small increase followed a decline and saturation, but the general level of coverage is 15--20 percentage points lower. This is the trade off for achieving the higher accuracy.

\paragraph{Alternative methods} To further complement the analysis, we use the same   experimental setup with the Markov chain model, this time on data from the French market. Here there results are less clear: top-4 accuracy on validation data is 0.15 when transitions are estimated on 10\% of training data, and stays around 0.16$\pm$0.005 when 20\%--100\% of the training data is used. Similarly, a rather constant trend around 0.25 is observed for catalog coverage. This is probably due to the fact that only one state, a single item,  is used to predict the next item in the sequence (this is the Markov property) according to transition probabilities. Estimation of these, from frequencies of pairwise consecutive %occurrences of 
items in training data, is relatively stable from the point of using 10\%--20\% of data. Adding  more  does not have a large effect on estimated parameters, neither will it help the model to predict  more complex sequential patterns.

% Finally, we also consider a couple of `classical' collaborative methods for making recommendations with implicit data: regularised matrix factorisation with alternating least squares \cite{hu2008collaborative} and  matrix factorisation with BPR \cite{rendle2012bpr}.

\begin{figure}[ht!]
    \centering
    \includegraphics[width=\linewidth]{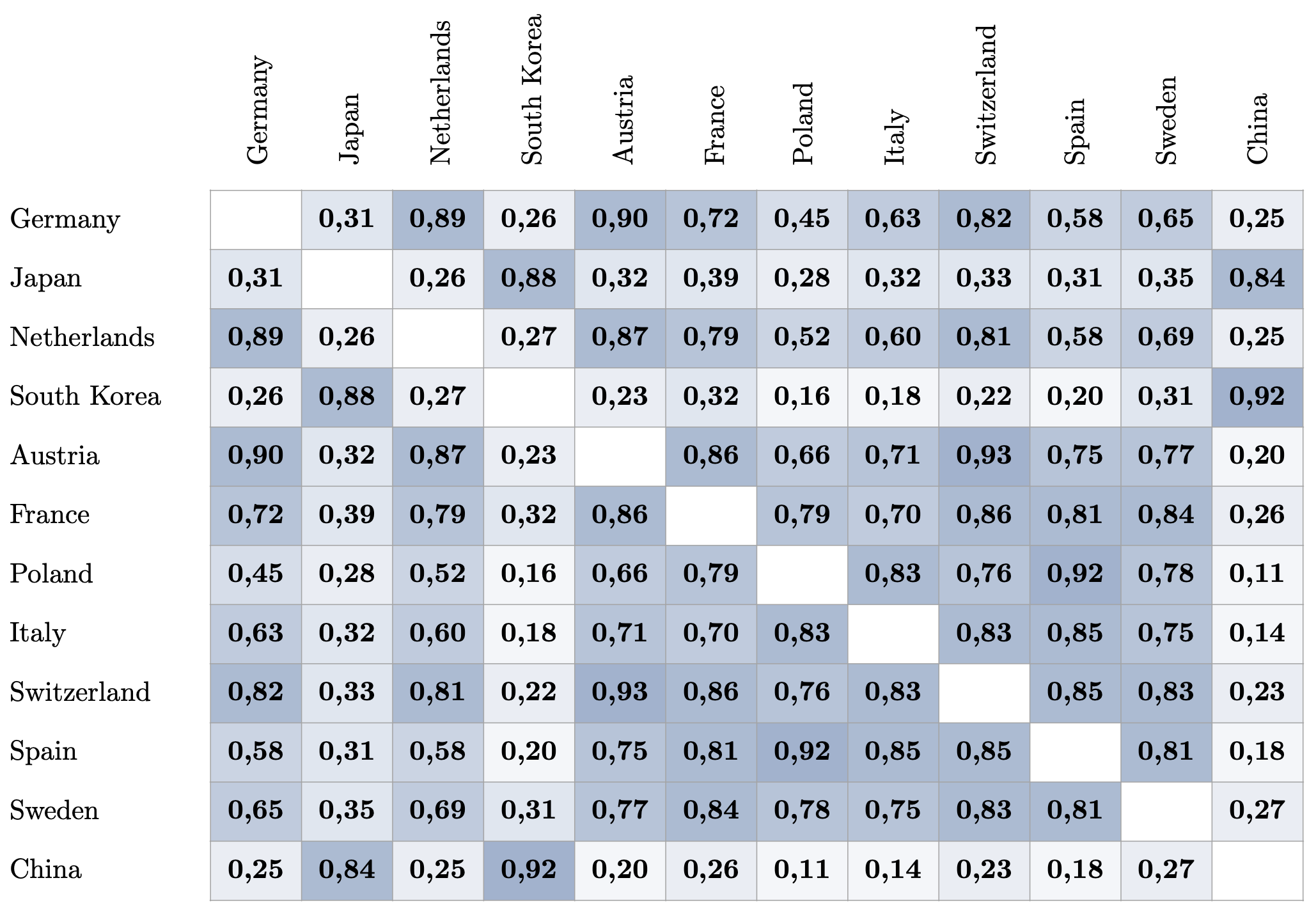}
    \caption{Similarities in purchase behaviour between different markets according to the measure based on an embedding learned by the neural network.}
    \label{fig:sim}
\end{figure}

%--
\subsection{Cross-market knowledge transfer}
%--
In this experiment, we first use the similarly measure of  Section \ref{sec:sim} to compare purchase behaviours of 12 markets from data of the online store. We then hypothesise  if a model trained with a similar market performs better than a model trained to a dissimilar market.

Figure \ref{fig:sim} contains computed similarities between the markets. %using the purchase embedding method. 
We note that there is a clear correlation between geographical closeness and  similarity measures. For example, the most similar countries to Germany in terms of purchase behaviour are Austria and Netherlands, while the most similar markets to China are South Korea and Japan. It is also possible to see that European markets are similar to each other while dissimilar to Asian markets. 
\begin{figure}[h!]
\begin{subfigure}{.42\textwidth}
  \centering
  \includegraphics[width=\linewidth]{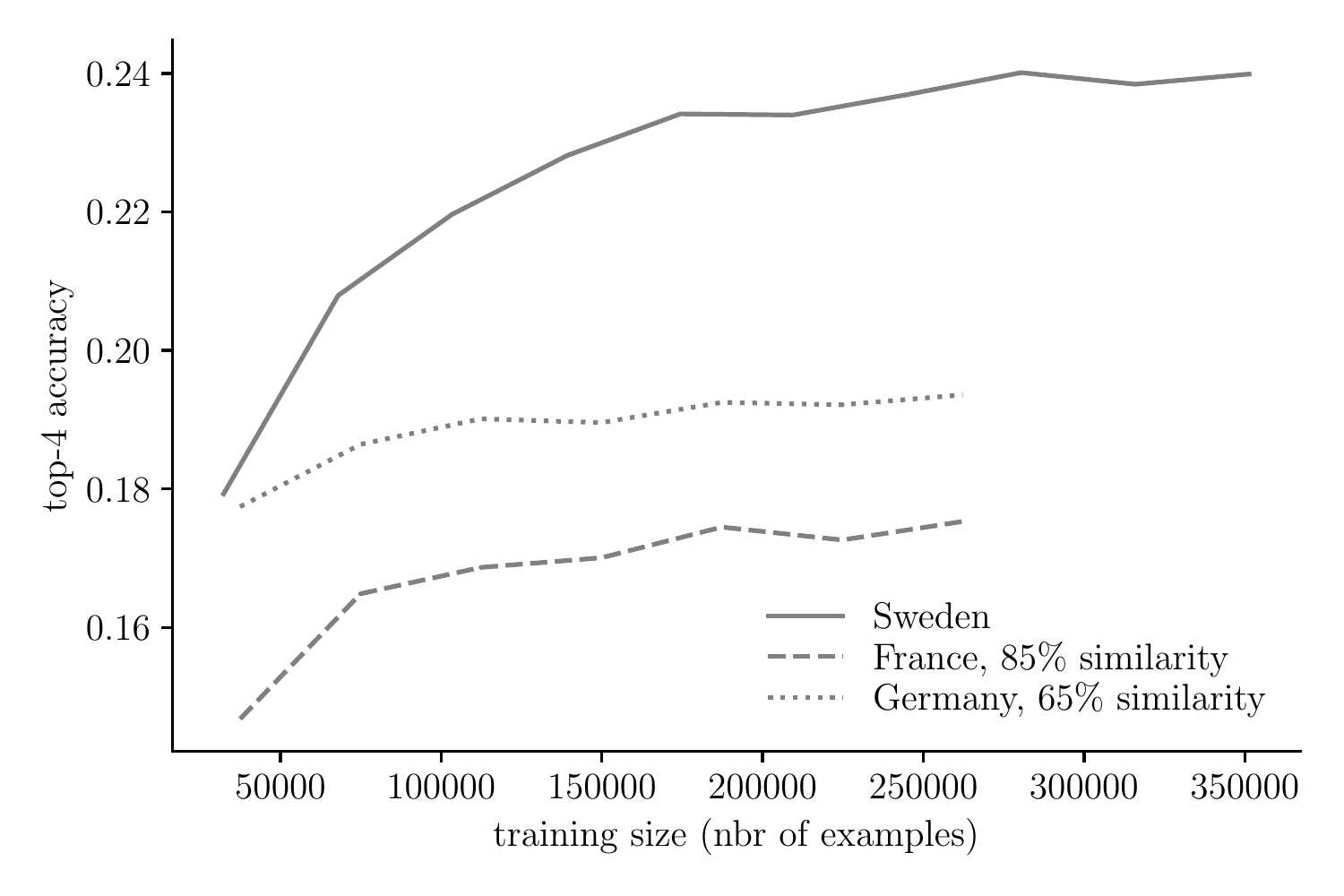}  
  \caption{}
  \label{fig:3-pre_at_k_swe}
\end{subfigure}
\begin{subfigure}{.42\textwidth}
  \centering
  \includegraphics[width=\linewidth]{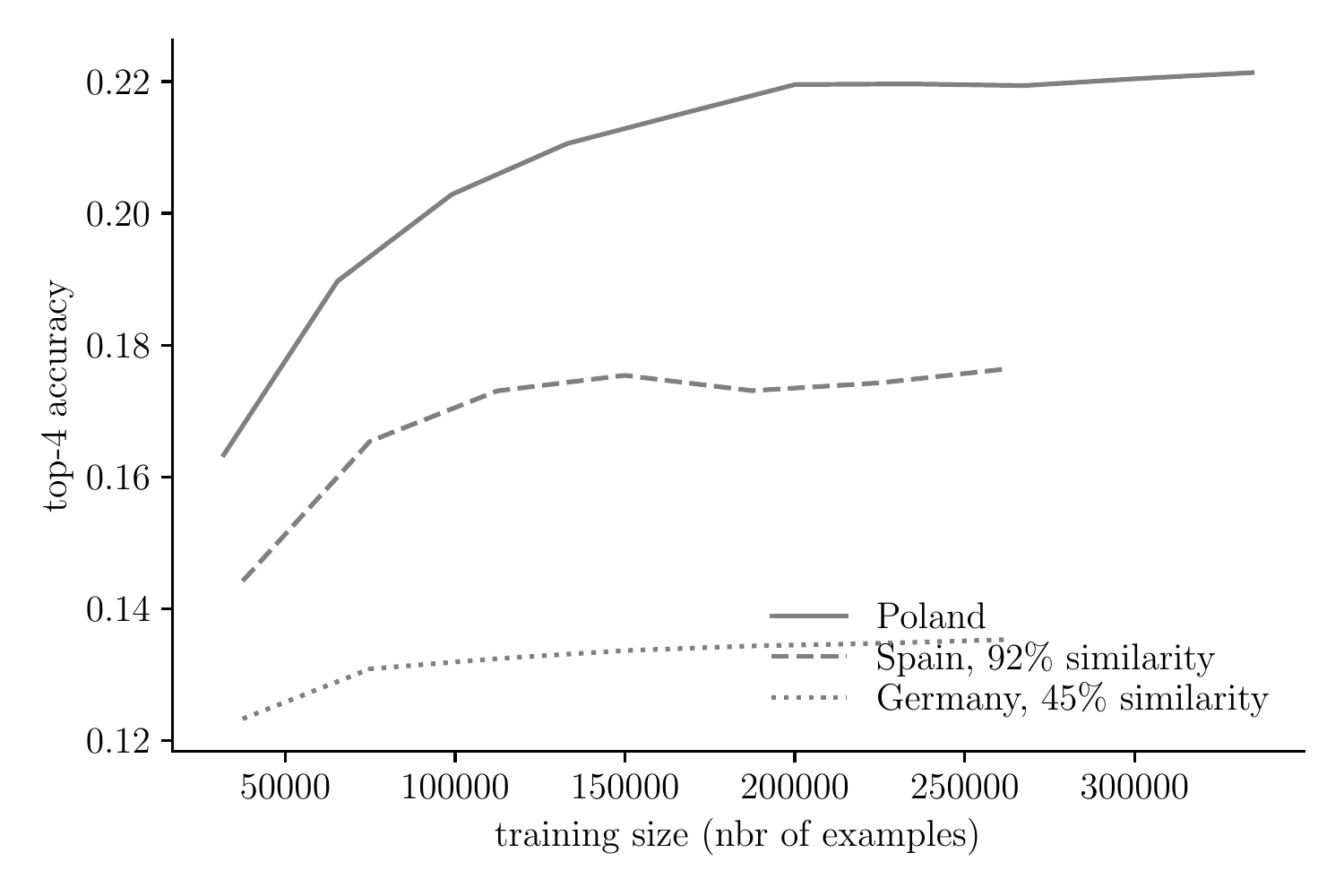}  
  \caption{}
  \label{fig:3-pre_at_k_pol}
\end{subfigure}
\begin{subfigure}{.42\textwidth}
  \centering
  \includegraphics[width=\linewidth]{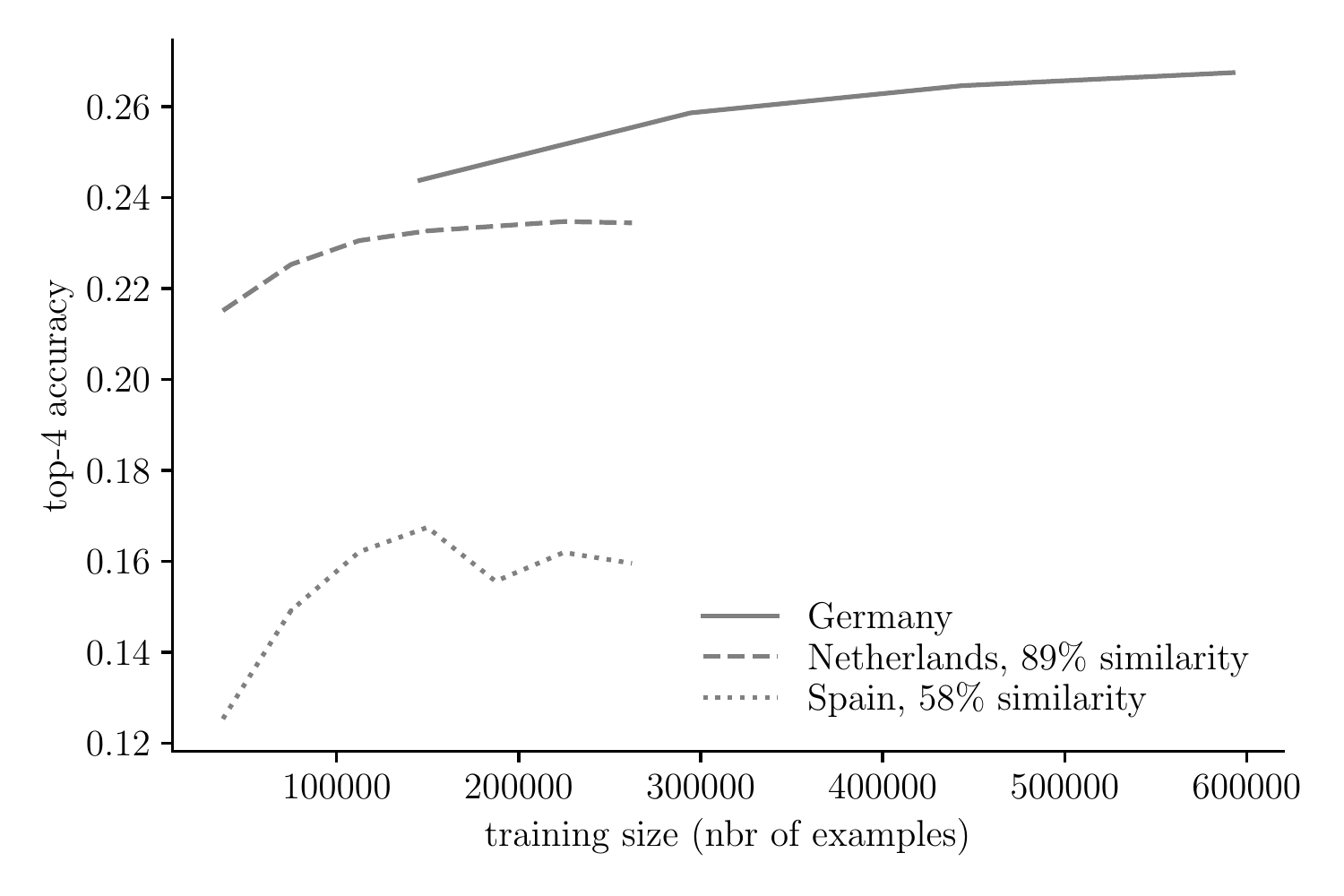}  
  \caption{}
  \label{fig:3-pre_at_k_de}
\end{subfigure}
\caption{Top-4 accuracy plotted as a function of dataset size when testing how well models trained on data from a secondary market performs on a local validation dataset.}
\label{fig:3-pre_at_k}
\end{figure}

%--
Our hypothesis is that similarities in purchase behaviour could be used do judge if data from one market could be leveraged for training a recommender system in another market.
We conduct an empirical study for this purpose. We construct a test to imitate a situation where one small market (here denoted \textbf{A}) lacks enough data to reach sufficient performance levels. To address the shortage, data from another \textit{similar} market is  added to the training dataset, to enable training with sufficient  data % reach the necessary amount to train a model which reaches 
to reach desirable performance levels on a validation set. We use top-K accuracy to evaluate  this experiment, as it is the metric %only metric, among the ones we have chosen to consider in this paper, 
that gives information about how well the recommender system  predicts purchase behaviour in a market. 

%--

%We test if our proposed similarity metric can be used for the selection of market that is suitable for transfer learning. 
To test if our similarity measure is suitable for selection, we %do this by training 
train three models on data from three different markets, denoted \textbf{A}, \textbf{B} and \textbf{C}. We then evaluate all three models using a validation dataset from market \textbf{A}. The markets are selected so that \textbf{A} and \textbf{B} have high similarity ($>0.85$) according to our %proposed similarity metric 
measure, and  \textbf{A} and \textbf{C} are dissimilar ($<0.65$). %according to the same metric. 
If the similarity measure performs as expected, the model trained on data from market \textbf{B} should outperform the model trained on data from market \textbf{C} when validated on  %validation set from 
market \textbf{A}. We conduct the test for three different triplets of markets. The results %from last these tests 
are presented in Figure \ref{fig:3-pre_at_k}.

As baseline, the model is trained and validated on data from the same market. This is the same setting as in Section \ref{sec:exp1} with results shown in Figure \ref{img:1pre@k}. We use the baseline to  judge how well the models with data from one market and then validated on another perform, compared to using data from the same market where it is validated. 

%--
In figure \ref{fig:3-pre_at_k_pol} and \ref{fig:3-pre_at_k_de} the model trained on data from a similar markets according to the suggested  measure, \textbf{B}, achieves higher Top-K accuracy than the model trained on data from a dissimilar market \textbf{C}. %as was predicted by the proposed similarity metric. 
However, in figure \ref{fig:3-pre_at_k_swe} the model trained on data from the dissimilar market, \textbf{C},  achieves a slightly higher Top-K accuracy than the model trained on data from the similar market.

%--
It is worth noting that there is a significant difference in achieved top-K accuracy between the baseline and the models trained on data from other markets, for all experiments reported in Figure \ref{fig:3-pre_at_k}. The exception is the case when data from the Netherlands has been used to train a model which has been validated on data from Germany, where top-K accuracy levels are very similar. 

%--
In all three examples in figure \ref{fig:3-pre_at_k} there is a significant difference in the top-K accuracy levels between the two markets, confirming that from which secondary market data is used is indeed %the training data is extracted is 
important. 

\section{Summary and Conclusion}\label{sec:conclusion}

\paragraph{Performance as a function of dataset size} In this paper, we show that a recommender system experience evident saturating effects in validation performance when the size of the training dataset grows. We opted to use a combination of three performance metrics, to better estimate  quality of the user experience. The three metrics are top-K-accuracy, catalog coverage and novelty. 

The saturating effect is especially evident in top-K-accuracy which %asymptotically 
approaches a maximum level for the specific problem. %, given a particular model configuration. 
Furthermore, we observe different behaviours for catalog coverage depending on which dataset we investigate. On the different datasets from the online store, we observe a decrease in  coverage when the amount of training data increases, while on the \textit{yoochoose} dataset we observe a saturating behaviour similar to the top-K-accuracy. %for models trained on data from the online store datasets. 

All of our experiments, considering all metrics, indicate  saturating effects on validation data when the amount of training data is increased. Our results thus further confirm results in \cite{larson2017towards} and complement their study in two aspects: evaluating on a validation set to capture the system's \textit{generalisation performance}, and evaluating in both accuracy and diversity metrics that are purposefully designed to capture \textit{good recommendations for the user}. %As \cite{larson2017towards} 
%Based on this, we make a case for working towards minimal necessary data when developing recommender systems. 

We find that there is an apparent trade-off between accuracy-focused metrics such as top-K-accuracy, and diversity-focused metrics such as catalog coverage and novelty, as indicated in \cite{div-acc-dilemma}. Hence, an optimal amount of data for training can be determined if a notion of optimality  is clearly defined: Depending on what is prioritized in the system design, the optimal amount of data varies. If accuracy is of high importance more data could lead to improvements. But if catalog coverage and novelty are more important, less data could be used without sacrificing too much accuracy. 

From these results, we conclude that %the apparent habit of \textit{data-greed} within the industry might be uncalled-for and that 
sufficient validation performance can be achieved without striving towards gathering as much data as possible. These results also give room for higher considerations of user privacy. Corporations could be more complaint towards users who are hesitant of sharing their personal data for privacy concerns, without risking significant performance losses within recommender systems.

\paragraph{Similarity measure and knowledge transfer} 
We propose a method for constructing a similarity measure, to measure compatibility between datasets in terms of purchase behaviour. %using embeddings. 
First, our tests  %results in a test of the  similarity measure 
confirm that it is possible to use data from a secondary market to complement a smaller dataset. %However, the market from which the data is selected is of great importance. The 
Second, we show that validation performance depends on \textit{which} market the data is taken from, and that performance varies significantly. %to a great extent. 
In none of our experiments, the models trained on data from another market were able to perform better than the model trained on data from the original market. However, if we trained to a `similar' market, the performance was generally better than when training to a `dissimilar' market. %This confirms that using a validation set and training set from the same market is preferred.

Our proposed metric  shows some promise in the sense that it manages to capture how geographical information correlates with purchase behaviour. For instance, it predicts that Germany, Netherlands, Switzerland and Austria have similar  behaviour. The metric also successfully predicted which data would be most successful for knowledge transfer in two out of the three tested cases. We find these results interesting and promising for future research. %but conclude that more work is needed on this topic. 

\paragraph{Conclusion}
Similar to \cite{larson2017towards} we conclude that it is not sensible to gather ever increasing amounts of data for training recommender systems. When giving the user a choice of sharing their personal data for the purpose of personalization, some users will inevitably opt out. Our results show that this decrease in data collected does not necessarily have to be a disadvantage. 
%First, 
We have shown that due to saturation effects in performance there is an amount of data that is \textit{enough} to reach sufficient performance levels. 
Secondly, we have shown that if there are not enough data available to reach  sufficient performance levels it is possible to use data from other similar domains to complete small datasets.

Finally, we propose a metric to judge what domains that are similar and where datasets are compatible to a higher extent. The results are promising and point us in a direction to further research on how to efficiently exploit such secondary data with transfer learning methods. %  but more research is needed on this topic.

With these results, we believe that there is a strong case for working towards minimal necessary data within recommender systems. This has considerable privacy benefits without necessarily having unfavourable effects on performance.

\subsection*{Acknowledgments}

We thank Kim Falk for helpful discussions.

\bibliographystyle{abbrv}
\bibliography{sample-base}

\end{document}